\documentclass{cpbtex}
\usepackage{longtable,booktabs}
\usepackage{graphicx}
\usepackage{dcolumn}
\usepackage{bm}
\usepackage{color}
\usepackage{upgreek}
\usepackage{threeparttable}
\usepackage[numbers,sort&compress]{natbib}
\usepackage{multirow}
\usepackage{epstopdf}
\usepackage{array}
\usepackage{bm}
\usepackage{CJK}

\begin{document}

\begin{CJK*}{UTF8}{gbsn}
	
\title{The effects of electron correlation and the Breit interaction on one- and two-electron one-photon transitions in double K hole states of He-like ions($10\!\leq\!Z\!\leq\!47$)\thanks{Project supported by the National Natural Science Foundation of China (Grant No.U1832126, 11874051).}}

% Title should be concise; avoid abbreviations if possible; and not begin with `A', `An', `The', or `Study on'.

\author{Xiaobin Ding(丁晓彬)$^{1}$\thanks{Corresponding author. E-mail:dingxb@nwnu.edu.cn}, Cunqiang Wu(吴存强)$^{1}$, Mingxin Cao(曹铭欣)$^{1}$, Denghong Zhang(张登红)$^{1}$, \\ Mingwu Zhang(张明武)$^{2}$, Yingli Xue(薛迎利)$^{2}$, Deyang Yu(于得洋)$^{2}$, Chengzhong Dong(董晨钟)$^{1}$\\
$^{1}${Key Laboratory of Atomic and Molecular Physics and Functional Materials}\\
{of Gansu Province, College of Physics and Electronic Engineering,}\\{ Northwest Normal University, Lanzhou 730070, China}\\  % The line break was forced via \\
$^{2}${Institute of Modern Physics, Chinese Academy of Sciences, Lanzhou 730000, China}} % The line break was forced via \\
%$^{3}${Third affiliation}}   % The line break was forced via \\

% 1. For Chinese authors, the name in Chinese characters should also be given. For example, Gang Liu(Áõ¸Õ), Xiao-Ming Li(ÀîÏþÃ÷)
% 2. Please ensure that every author approves the submission of the manuscript
% 3. Abbreviations should not be used in the affiliations

\date{\today}
\maketitle

\begin{abstract}
The X-ray energies and transition rates associated with single and double electron radiative transitions from the double K hole state $2s2p$ to the $1s2s$ and $1s^{2}$ configurations of 11 selected He-like ions ($10\!\leq\!Z\!\leq\!47$) are calculated using the fully relativistic multi-configuration Dirac-Fock method(MCDF). An appropriate electron correlation model is constructed with the aid of the active space method, which allows the electron correlation effects to be studied efficiently. The contributions of electron correlation and the Breit interaction to the transition properties are analyzed in detail. It is found that the two-electron one-photon (TEOP) transition is correlation sensitive. The Breit interaction and electron correlation both contribute significantly to the radiative transition properties of the double K hole state of He-like ions. Good agreement between the present calculation and previous work is achieved. The calculated data will be helpful to future investigations on double K hole decay processes of He-like ions.
\end{abstract}



\section{Introduction}
The energy level structures and radiative decay processes of inner-shell hole states are an important issue in atomic physics \cite{Hoogkamer1976,PhysRevLett.37.59,PhysRevLett.36.164,PhysRevLett.37.63,STOLLER,Safronova1977}. An inner-shell hole state occurs when the inner shell orbital of an atom or ion is unoccupied, while the outer shell orbital is occupied by electrons. Inner-shell hole states have been observed in high-energy ion-atom collisions\cite{Kumar2006,PhysRevA.44.1836,PhysRevA.67.042703}, synchrotron radiation\cite{PhysRevA.62.052519}, laser-produced plasmas\cite{Boiko1977}, ion beam-foil spectroscopy\cite{Andriamonje1991}, Tokamak\cite{PhysRevA.29.661}, and solar flares\cite{Phillips2004a}. They can also be produced by electron excitation or the ionization of the inner shells of atoms or ions\cite{Lopez-Urrutia2005}, as well as in inner-shell photoionization or photoexcitation processes with high-energy photons \cite{PhysRevA.79.032708}. These exotic atoms are extremely unstable and mainly decay through non-radiative Auger processes\cite{PhysRevA.44.239,article,Inhester2012} and radiative processes. The former processes are usually more efficient than the latter. With the development of X-ray spectroscopy, weak signal detection technology has helped scientists to understand such processes from the photon perspective.

It is also possible to create an ion with an empty innermost shell, forming a double K shell hole state\cite{PhysRevLett.102.073006}. Generally, the radiative de-excitation of an atom with an initially empty K shell may take place either through the more probable one-electron one-photon (OEOP) transition or through the competing weak two-electron one-photon (TEOP) transition. The initially double K hole state $2s2p$ in He-like ions can decay either through an OEOP transition to a single excited state $1s2s$, where a $2p$ electron transitioning to $1s$ with a spectator $2s$ electron, or through a TEOP transition, in which both electrons in the $2s$ and $2p$ orbitals transition to $1s$ orbitals simultaneously, producing the ground state $1s^{2}$ due to electron correlation effects. The TEOP process was first predicted theoretically by Heisenberg in 1925\cite{Heisenberg1925} and was observed by W\"{o}lfli \emph{et al.} in ion-atom collision experiments between Ni-Ni, Ni-Fe, Fe-Ni, and Fe-Fe in 1975\cite{PhysRevLett.35.656}. Since then, TEOP transitions have been widely studied both theoretically and experimentally\cite{Porquet2010,0004-637X-482-2-1076,PhysRevA.75.062502,PhysRevA.88.052522,PhysRevA.93.032516,PhysRevA.15.154,PhysRevA.84.062506,PhysRevA.88.012501,PhysRevA.90.032509,Elton2000,Trabert1982,PhysRevA.59.245,Tawara2002b}.

The TEOP process is forbidden in the independent particle approximation of an atom. Investigations of this process are helpful for explaining electron correlation effects, relativistic effects and quantum electro-dynamics (QED) effects on the energy level structure and radiative transitions of these exotic atoms. Insights into the electron coupling of complex atom systems are also helpful. For astrophysical and laboratory plasmas, some important diagnostics information regarding the composition, temperature, and density has also been provided by these basic atomic physics processes\cite{Porquet2010,0004-637X-482-2-1076}.

There have been many works related to the energy levels and transition properties of inner-shell hole states in the past several decades\cite{PhysRevA.75.062502,PhysRevA.88.052522,PhysRevA.93.032516,PhysRevA.15.154,PhysRevA.84.062506,PhysRevA.88.012501,PhysRevA.90.032509,Elton2000,Trabert1982,PhysRevA.59.245,Tawara2002b,Kadrekar2010a,Li2010}, but only a few studies have focused on He-like ions\cite{PhysRevA.15.154,PhysRevA.84.062506,PhysRevA.88.012501,PhysRevA.90.032509,Elton2000,Trabert1982,PhysRevA.59.245,Tawara2002b}. The He-like ion is a two-electron system with simple structure and electron correlation effect and is a good candidate with which to study the TEOP process. Kadrekar and Natarajan calculated the transition properties and branching ratios between OEOP and TEOP transitions in He-like ions with $2s2p$ configurations using the multi-configuration Dirac-Fock (MCDF) method \cite{PhysRevA.84.062506} and found that the contribution from the TEOP transition is considerable for low-Z ions. The influence of the configuration interaction on single-electron allowed E1 transitions is negligible. They also calculated both OEOP and TEOP transition rates from $2s2p$ and $2p^{2}$ of He-like Ni, including electric dipole transitions (E1) and magnetic quadrupole transitions (M2) \cite{PhysRevA.88.012501} and found that higher order corrections are more important for $\Delta n = 0$ than for $\Delta n = 1$ transitions of He-like Ni. After that, Natarajan conducted research on the orthogonality of the basis. The biorthogonal and common basis sets give almost the same transition rates for light and medium heavy elements while the differences are substantial for heavy elements\cite{PhysRevA.90.032509}. The contributions from correlation and higher-order corrections, consisting of Breit and QED effects, to the energies and transition rates were analyzed. Experimentally, transitions from $2s^{2}$-$1s2p$ in He-like Si have been observed in laser-produced plasma experiments at the TRIDENT facility by Elton \emph{et al.} \cite{Elton2000}. Tawara and Richard \emph{et al.} have observed Ar K X-rays under 60 keV/u Ar$^{16+}$-Ar collisions from the KSU EBIS \cite{Tawara2002b}.

Previous theoretical and experimental investigations of OEOP and TEOP transitions have mostly focused on the low-Z atoms, with only a few works focusing on high-Z ions \cite{Kadrekar2010a,Li2010}. The present work provides an MCDF calculation of OEOP and TEOP transitions from double K hole $2s2p$ configurations in 11 selected He-like ions ($10\!\leq\!Z\!\leq\!47$). The electron correlation effects are accounted for by choosing appropriate electron correlation models using the active space method. The Breit interaction and QED effects are included perturbatively in relativistic configuration interaction (RCI) calculations. The finite nuclear size effects are described by a two-parameter Fermi distribution model. The purpose of the present calculations is to explore how the effects of electron correlation and the Breit interaction on the transition energies and rates of OEOP and TEOP transitions vary with increasing Z. The results will be helpful to future theoretical and experimental work on the radiative decay processes of double K hole states. The calculations were performed using the Grasp2K code \cite{Joensson2007}.

\section{Theory}\label{sec2}
The multi-configuration Dirac-Fock (MCDF) method has been widely used to investigate relativistic, electron correlation, Breit interaction, and quantum electrodynamics (QED) effects on the structure and transitions of complex atoms or ions based on relativistic atomic theory \cite{Ding2011a,Ding,Aggarwal,Liu2017,Hu2011}. The method was expounded in Grant's monograph \cite{Grant2007} and implemented in the Grasp family code \cite{Grant1980,Desclaux,K.1989,GRASP92,Joensson2007}. Here, only a brief introduction to the MCDF method is provided.

In the MCDF method, the atomic state wave function (ASFs) $\Psi$($PJM_{J}$) for a given state with certain parity \textit{P}, total angular momentum \textit{J}, and its \textit{z} component $M_{J}$ is represented by a linear combination of configuration state functions (CSFs) $\Phi$($\gamma$$_{i}$$PJM_{J}$) with the same \textit{P}, \textit{J}, $M_{J}$, which can be expressed as:

\begin{eqnarray}
\Psi(PJM_{J})=\sum_{i=1}^{N_{c}}c_{i}\Phi(\gamma_{i}PJM_{J}).
\end{eqnarray}

where $N_{c}$ is the number of CSFs and $\gamma_{i}$ denotes all the other quantum numbers necessary to define the configuration, $c_{i}$ is the mixing coefficient. The CSFs are linear combinations of the Slater determinants of the many-particle system consisting of single electron orbital wave functions. The extended optimal level (EOL) scheme is used in the self-consistent field (SCF) calculation to optimize the radial wave functions. The mixing coefficients $c_{i}$ of CSFs are determined variationally by optimizing the energy expectation value of the Dirac-Coulomb Hamiltonian, which is defined as in the following equation:

\begin{equation}
H_{DC}=\sum_{i=1}^{N}[c\alpha_{i}\cdot p_{i}+(\beta_{i}-1)c^{2}+V_{i}^{N}]+\sum_{i>j}^{N}\frac{1}{r_{ij}}
\end{equation}

In order to include the higher-order interaction such as Breit interaction and QED effects, the RCI calculation with the same CSFs as SCF calculation was done. The transverse photon interaction plays a dominant role in the calculations, especially for high-Z ions, which can be expressed as follows:

\begin{equation}\label{breit}
H_{trans}=\sum_{i,j}^{N}[\frac{\alpha_{i}\cdot p_{i}cos(\omega_{ij})}{r_{ij}}+(\alpha_{i}\cdot\bigtriangledown_i)(\alpha_{j}\cdot\bigtriangledown_j)\frac{cos(\omega_{ij})-1}{\omega_{ij}^{2}r_{ij}}]
\end{equation}

The Breit interaction is the low-frequency limit of eq (\ref{breit}). QED effects including vacuum polarization and self-energy are also taken into account in the present calculation perturbatively.

\section{Electron correlation model and calculation strategy}\label{sec3}

The electron correlation effects are taken into account by choosing an appropriate electron correlation model. The correlation model used in the present calculation is similar to the model used by Kadrekar and Natarajan \cite{PhysRevA.84.062506}. The major electron correlation effects can be captured by including the CSFs, which were formed by allowing single and double (SD) excitations from the interested reference configurations to some virtual orbital space. The configuration space was extended by increasing the active orbital set layer by layer to study the correlation contributions. Generally, the zero-order Dirac-Fock (DF) wave functions were first generated from the reference configurations of He-like ions in EOL mode for the initial and final states. In the EOL method, the radial wave functions and the mixing coefficients are determined by optimizing the energy functional, which is the weighted sum of the selected eigenstates. For a double K hole state, the minimum basis (MB) was generated by considering limited expansion and allowing SD substitutions of electrons from the reference configurations. Since this procedure results in better optimized wave functions than the DF functions, all the examinations of the correlation effects here are carried out with respect to the MB. Then, the active space was expanded to the first layer, i.e. $n=3, l=2$ (\{\textit{n}3\textit{l}2\}) virtual orbitals and all the newly added orbital functions were optimized while the $1s$, $2s$, and $2p$ orbitals were kept fixed from the MB. These steps were repeated, increasing the virtual orbitals to ensure that the eigenenergy and wave function converged. To ensure the stability of the numeric data and reduce the calculation time, only the newly added layer was optimized at each step and the previously calculated orbits were all kept frozen. As the virtual orbitals increased, the number of CSFs increased rapidly. In this method, the electrons from the occupied orbitals are excited to unoccupied orbitals in the active space. Since the orbitals with the same principal quantum number n have similar energies, the active set is expanded in layers of n and the \{nl\} set includes all the orbitals with l=0 to n-1. However, our calculations show that higher l values contribute very little. So, the present work was restricted to n=1 to 6 and l=0 to 3 to keep the calculation traceable and manageable.

\section{Results and discussion}\label{Results}

The energy levels and transition properties of the He-like Ne, Si, Ar, Ca, Fe, Ni, Cu, Zn, Kr, Nb, and Ag ions were calculated using MCDF with the active space method. The energy levels (in eV) of the double excited configuration $2s2p$ and the single excited configuration $1s2s$ of He-like Ne and Ag ions are presented in Table~\ref{Tab1} to show the convergence. Since the correlation model of MB provides better optimized wave functions than the DF functions, all our investigations on the correlation effects and higher-order corrections were carried out with respect to the MB. It can be speculated from the table that with an increase in the active space, the eigenenergies tend to converge for both low-Z and high-Z ions. The energy $E$ of $2s2p$ relative to the ground state $^1S_0 \, 1s^{2}$ of He-like Ne was provided with available theoretical results. Excellent agreement of the relative errors ($\leq0.1\%$) between the present calculation and previous work that also used the MCDF method was achieved. Therefore the present calculation was restricted to the $\{n6l3\}$ correlation models.

\begin{table}
\centering
\caption {Energies (in eV) of the initial and final states of He-like Ne and Ag ions in various active space sets. The notation DF denotes the Dirac-Fock calculation, MB the minimum basis, \{$n$a$l$b\} the active set consisting of all orbitals from $n$=a to $l$=b, and $E$ the energy relative to the ground state $^1S_0$ $1s^{2}$. For details see Sec. \ref{sec3}. }
\label{Tab1}
\setlength{\tabcolsep}{4.0mm}{
%\scriptsize
\begin{tabular}{ccccccccc}
\midrule
\hline
\multicolumn{7}{c}{He-like Ne}                                                                \\\hline
			\multirow{2}{*}{Active sets}  & \multicolumn{4}{c}{2s2p} &   \multicolumn{2}{c}{1s2s} \\
			& $^{3}$P$_{0}$ & $^{1}$P$_{1}$ & $^{3}$P$_{2}$ & $^{3}$P$_{1}$& $^{3}$S$_{1}$ & $^{1}$S$_{0}$\\
			DF	    & -645.71     & -629.80     & -645.30     & -645.58  & -1653.06    & -1643.25   \\	
			MB 	    & -645.49     & -629.38     & -645.08     & -645.36  & -1652.95    & -1641.73    \\	
			n3l2     & -645.68     & -630.16     & -645.26     & -645.55  & -1653.00    & -1642.07     \\	
			n4l3     & -645.72     & -630.36     & -645.31 	  & -645.59  & -1653.01    & -1642.16    \\	
			n5l3     & -645.74     & -630.74     & -645.38     & -645.62  & -1653.03    & -1642.23   \\	
			n6l3     & -645.80     & -630.92     & -645.39     & -645.67  & -1653.03    & -1642.26    \\
			\\
			E      &  1911.70    &  1926.58    &  1912.11    &  1911.83 &  904.47     &  915.24    \\
			Ref$^a$  &  1911.48    &  1926.13    &  1911.89    &  1911.60 &  904.41     &  914.82    \\
			NIST     &  1912.26    &  1926.63    &  1912.83    &  1911.97 &  905.08     &  915.34    \\		
			~\\
			
			\multicolumn{7}{c}{He-like Ag}                                                                \\\hline
			\multirow{2}{*}{Active sets}  & \multicolumn{4}{c}{2s2p}&\multicolumn{2}{c}{1s2s}\cr
			& $^{3}$P$_{0}$& $^{1}$P$_{1}$& $^{3}$P$_{2}$& $^{3}$P$_{1}$& $^{3}$S$_{1}$& $^{1}$S$_{0}$\\
			DF	    & -15437.72   & -15146.59   &	-15203.39   & -15417.21   & -38550.42   & -38489.71 \\	
			MB	    & -15437.46   & -15146.22   &	-15203.17   & -15416.90   & -38550.33   & -38487.92 \\	
			n3l2	    & -15437.65   & -15146.84   & -15203.36     & -15417.21   & -38550.38   & -38488.33\\	
			n4l3  	& -15437.70   & -15147.04   &	-15203.41   & -15417.29	  & -38550.39   & -38488.42 \\
			n5l3	    & -15437.72   & -15147.42   &	-15203.48   & -15417.35   & -38550.41   & -38488.51\\
			n6l3	    & -15437.78   & -15147.53   &	-15203.50   & -15417.48   & -38550.41   & -38488.54 \\
			\hline
			\hline
	\end{tabular}}
	\begin{tablenotes}
		\item[] $^{\rm a}$ Reference\cite{PhysRevA.84.062506}
	\end{tablenotes}
\end{table}

The transition energies in eV of the OEOP transitions from the $2s2p$ configuration to the $1s2s$ configuration of He-like ions ($10\!\leq\!Z\!\leq\!47$) are presented in Table~\ref{Tab2}. The results for $Z\!\leq\!26$ He-like ions agree well with the available experimental data and other theoretical calculation results. The average relative error of the current calculation compared to the experimental observation is about $0.01\%$-$0.09\%$. Results for the ions with $Z\geq\!28$ were also calculated for the present work. To the best of the authors' knowledge, corresponding experimental and theoretical data was otherwise unavailable. Therefore, it will now be helpful to future experimental and theoretical investigations.

\begin{table}
\footnotesize\rm
\centering
\caption{Transition energies (in eV) of one-electron radiative transitions from $2s2p$ configuration in He-like ions. '*' denotes the spin-forbidden transition. }
\label{Tab2}
\setlength{\tabcolsep}{4.3mm}{
%\scriptsize
\begin{tabular}{cccccccc}
\hline
\hline
\multicolumn{8}{c}{Energy}                                                               \\\hline
\multirow{1}{*}{Z}
           &       & $^{3}$P$_{1}$-$^{1}$S$_{0}$$^*$ & $^{3}$P$_{0}$-$^{3}$S$_{1}$ &  $^{3}$P$_{1}$- $^{3}$S$_{1}$ & $^{3}$P$_{2}$-$^{3}$S$_{1}$& $^{1}$P$_{1}$-$^{1}$S$_{0}$ &  $^{1}$P$_{1}$-$^{3}$S$_{1}$$^*$\\
           \hline
   10	 &            & 996.38   & 1007.09    & 1007.22	    & 1007.50    & 1011.14       & 1021.97  \\	
    	 & Ref$^a$    & 996.79   & 1007.07    & 1007.20	    & 1007.48    & 1011.32       & 1021.73 \\	
         & Expt.      &          &            & 1007.86$^b$ &            &               &           \\
         & Theory$^b$ & 996.92   & 1007.0     & 1007.2      & 1007.5     & 1011.2        & 1021.4   \\
   14    &            & 1968.59  & 1983.91    & 1984.41     & 1985.54    & 1990.22       & 2006.04  \\
         & Ref$^a$    & 1968.97  & 1983.88    & 1984.39     & 1985.52    & 1990.39       & 2005.80  \\
         & Expt.      &          &            &             & 1985.8$^c$ & 1991.7$^c$    &                        \\
         & Theory$^b$ & 1969.1   & 1983.9     & 1984.4      & 1985.5     & 1990.3        & 2005.5  \\
   18    &            & 3272.06	 & 3291.73	  & 3293.05	    & 3296.28    & 3301.40       & 3322.40 \\
   	     & Ref$^a$    & 3272.44	 & 3291.69	  & 3293.02	    & 3296.26    & 3301.56       & 3322.15 \\
         & Theory$^b$ & 3272.6   & 3291.7     & 3293.0      & 3296.2     & 3301.5        & 3321.9 \\
   20    &            & 4048.97	 & 4070.68    & 4072.63	    & 4077.67    & 4082.73       & 4106.39 \\
         & Ref$^a$    & 4049.34  & 4070.64    & 4072.60	    & 4077.66	 & 4082.89       & 4106.15\\
         & Theory$^b$ & 4049.5   & 4070.6     & 4072.6      & 4077.6     & 4082.8        & 4105.9   \\
   26    &            & 6886.05	 & 6913.88	  & 6918.15	    & 6933.78	 & 6937.35       & 6969.45       \\
         & Ref$^a$    & 6886.41	 & 6913.32	  & 6918.10	    & 6933.77	 & 6937.52       & 6969.20      \\
         & Expt.      &          &            & 6910$^d$    &            & 6942$^e$      &              \\
         & Theory$^b$ & 6886.7   & 6913.4     & 6918.1      & 6933.8     & 6937.6        & 6969.0     \\
   28    &            & 8002.38  & 8031.47	  & 8037.45	    & 8059.10	 & 8061.75       & 8096.83   \\
   29    &            & 8592.88  & 8622.86    & 8629.48	    & 8654.75	 & 8656.88       & 8693.48 \\
   30    &            & 9205.07	 & 9235.93	  & 9243.22	    & 9272.57	 & 9274.13       & 9312.28 \\
   36    &            & 13338.82 & 13375.35	  & 13386.87	& 13452.79   & 13450.21      & 13498.25\\
   41    &            & 17399.28 & 17441.38	  & 17456.47	& 17573.86   & 17567.01      & 17624.2\\
   47    &            & 23036.86 & 23086.18	  & 23105.36	& 23319.99   & 23307.07      & 23376.57\\
\hline
\hline
\end{tabular}}
\begin{tablenotes}
\item[] $^{\rm a}$ Reference\cite{PhysRevA.84.062506}.
\item[] $^{\rm b}$ Reference\cite{Goryaev2017}.
\item[] $^{\rm c}$ Reference\cite{Mosnier1986}.
\item[] $^{\rm d}$ Reference\cite{Nandi2008}.
\item[] $^{\rm e}$ Reference\cite{Phillips2004a}\\
\end{tablenotes}
\end{table}

In the calculation of transition properties in relativistic atomic theory, the Babushkin (B) and Coulomb (C) gauges are often used, which correspond to the length and velocity gauges in non-relativistic quantum mechanics, respectively. These are equivalent when exact wave functions are used, but they usually give rather different results when approximate wave functions are used. The consistency of the transition rates from different gauges therefore indicates the accuracy of the wave function to some extent. The ratio of the transition rates from the Babushkin and Coulomb gauges has often been adopted as a criterion for ensuring the accuracy of the wave function and the calculation results. In our calculations, the ratio of the transition rates from different correlation models tended towards 1.00 with increased active space. This indicates that the wave function used in the present calculation is good and that the most important correlation effects were included in the present work.

The transition rates of the OEOP transition from $2s2p$ to $1s2s$ of He-like ions ($10\!\leq\!Z\!\leq\!47$) are presented in Table~\ref{Tab3}. For brevity, only the transition rates in the Babushkin gauge are given in the table. The current calculated transition rates are in good agreement with the result calculated by Kadrekar and Natarajan using the MCDF method \cite{PhysRevA.84.062506} and by Goryaev \emph{et al.} using the Z-expansion method \cite{Goryaev2017}. The Z-expansion method is based on perturbation theory and a hydrogen-like basis, while MCDF includes electron correlation effectively. Four allowed transitions and two dipole forbidden transitions are listed in the table. For transitions from the same initial state $^{3}$P$_{1}$ to different final states $^{1}$S$_{0}$ and $^{3}$S$_{1}$, the ratio of the two transition rates is approximately $10^{-3}$ when Z=10, while the ratio increases to $10^{-1}$ when Z=47. This indicates that the intensity of these dipole forbidden transitions increases sharply with increasing Z, which provides a candidate for the observation of E1 forbidden transitions in high-Z ions. For high temperature plasma, some important diagnostics information is provided by these transitions.

\begin{table}
\footnotesize\rm
\centering
\caption{Transition rates (in sec$^{-1}$) of the one-electron radiative transitions from the $2s2p$ configuration in He-like ions, with '*' denoting the spin-forbidden transition.}
\label{Tab3}
\setlength{\tabcolsep}{4.0mm}{
%\scriptsize
\begin{tabular}{cccccccc}
\hline
\hline
\multicolumn{8}{c}{Rate}                                                                \\\hline
\multirow{1}{*}{Z}
           &       & $^{3}$P$_{1}$-$^{1}$S$_{0}$$^*$ &$^{3}$P$_{0}$-$^{3}$S$_{1}$ &$^{3}$P$_{1}$- $^{3}$S$_{1}$ &$^{3}$P$_{2}$-$^{3}$S$_{1}$& $^{1}$P$_{1}$-$^{1}$S$_{0}$ & $^{1}$P$_{1}$-$^{3}$S$_{1}$$^*$\\
\hline

   10	 &            &1.042(9)     &5.667(12)    & 5.661(12)    & 5.650(12)  & 5.661(12) & 8.316(8) \\	
    	 & Ref$^a$    &1.246(9)     &5.755(12)	  & 5.751(12)    & 5.744(12)  & 5.946(12) & 1.046(9)\\	
         & Theory$^b$ &1.17(9)      &5.79(12)     & 5.80(12)     & 5.80(12)   & 6.02(12)  & 1.22(9) \\
   14    &            &3.090(10)    &2.245(13)    & 2.239(13)    & 2.236(13)  & 2.237(13) & 2.736(10)\\
         & Ref$^a$    &3.375(10)    &2.268(13)    & 2.262(13)    & 2.259(13)  & 2.318(13) & 2.938(10)\\
         & Ref$^b$    &3.19(10)     &2.29(13)     & 2.28(13)     & 2.29(13)   & 2.35(13)  & 3.31(11)\\
   18    &            &3.611(11) 	&6.240(13)    & 6.179(13)    & 6.198(13)  & 6.178(13) & 3.318(11)\\
   	     & Ref$^a$    &3.818(11) 	&6.285(13)	  & 6.238(13)    & 6.246(13)  & 6.346(13) & 3.465(11)\\
         & Theory$^b$ &3.63(11)     &6.36(13)     & 6.33(13)     & 6.38(13)   & 6.47(13)  & 3.75(11)\\
   20    &            &9.8669(11)    &9.568(13)    & 9.432(13)    & 9.489(13)  & 9.419(13) & 9.153(11)\\
         & Ref$^a$    &1.034(12)    &9.628(13)    & 9.509(13)  	 & 9.554(13)  & 9.646(13) & 9.472(11)\\
         & Theory$^b$ &9.86(11)     &9.76(13)     & 9.67(13)     & 9.81(13)   & 9.87(13)  & 1.01(12) \\
   26    &            &1.067(13)    &2.768(14)	  & 2.655(14)    & 2.730(14)  & 2.636(14) & 1.005(13)	     \\
         & Ref$^a$    &1.100(13)    &2.780(14)	  & 2.666(14)    & 2.744(14)  & 2.683(14) & 1.025(13)     \\
         & Theory$^b$ &1.06(13)     &2.84(14)     & 2.74(14)     & 2.87(14)   & 2.79(14)  & 1.08(13)        \\
   28    &            &1.992(13)    &3.738(14) 	  & 3.576(14)    & 3.676(14)  & 3.496(14) & 1.881(13)              \\
   29    &            & 2.652(13)   &4.304(14)    & 4.029(14)  	 & 4.232(14)  & 3.990(14) & 2.505(13)\\
   30    &            & 3.472(13)   &4.936(14)    & 4.578(14)  	 & 4.847(14)  & 4.529(14) & 3.283(13)\\
   36    &            & 1.318(14)   &1.031(15)	  & 8.977(14)    & 1.005(15)  & 8.817(14) & 1.245(14)\\
   41    &            & 3.013(14)	&1.744(15)	  & 1.440(15)    & 1.686(15)  & 1.405(15) & 2.884(14)\\
   47    &            & 6.548(14)	&3.029(15)	  & 2.371(15)    & 2.897(15)  & 2.283(15) & 6.141(14)\\
\hline
\hline
\end{tabular}}
\begin{tablenotes}
\item[] $^{\rm a}$ Reference\cite{PhysRevA.84.062506}.
\item[] $^{\rm b}$ Reference\cite{Goryaev2017}.\\
\end{tablenotes}
\end{table}

The transition energies and rates of TEOP transitions from the initial $2s2p$ configuration to the final $1s^{2}$ configuration are listed in Table~\ref{Tab4}. The ratio of the transition rates in the Babushkin and Coulomb gauges is about $1.2$-$1.5$. The TEOP transition energy is approximately twice the corresponding OEOP transition energy, as expected. In general, good agreement between the present rate and the length gauge rate of Kadrekar\emph{et al.} \cite{PhysRevA.84.062506} can be obtained.

\begin{figure*}[!h]
\begin{minipage}[t]{1\linewidth}
\centering
\includegraphics[width=6in]{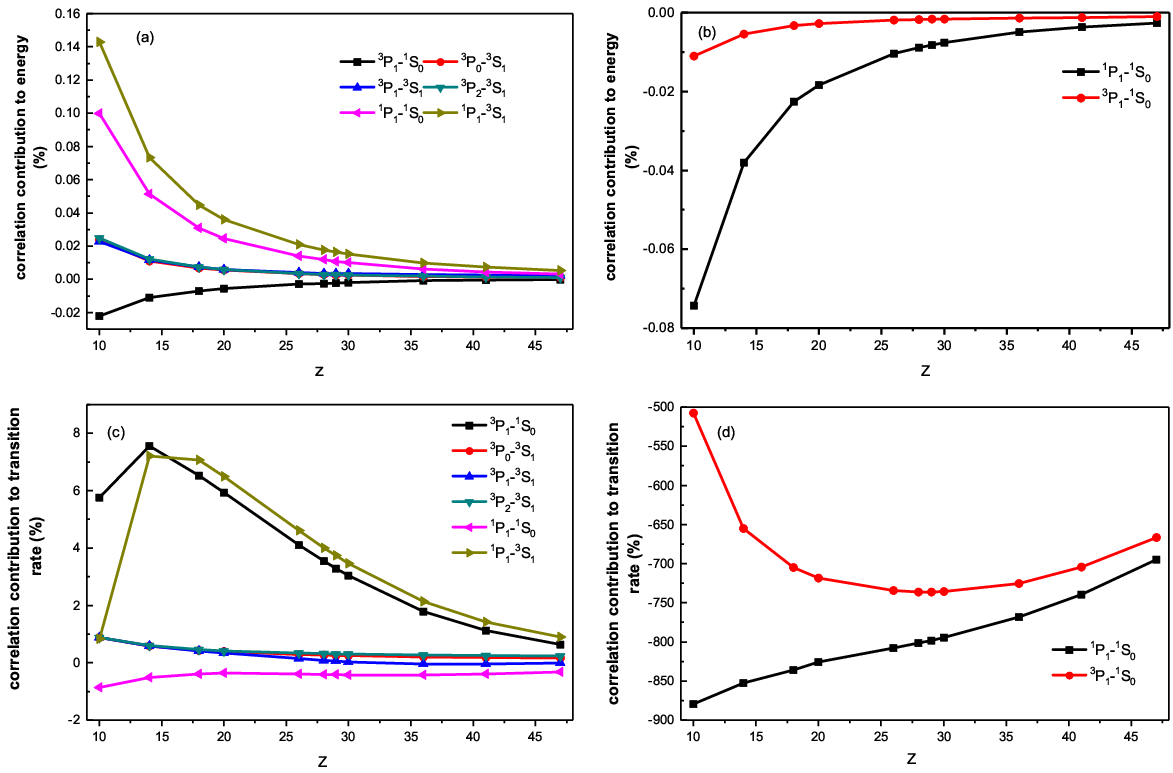}
%\captionsetup{font={scriptsize}}   %
\centering
\caption{The electron correlation effect on the transition energies and rates for OEOP and TEOP transitions in He-like ions. (a) The percentage correlation contribution to the OEOP transition energies of $2s2p-1s2s$. (b) The percentage contribution to the TEOP transition energies of $2s2p-1s^{2}$. (c) The percentage contribution to the OEOP transition rate in the length gauge of $2s2p-1s2s$. (d) The percentage contribution to the TEOP transition rate in the length gauge of $2s2p-1s^{2}$.\label{fig1}}
\end{minipage}
\end{figure*}

The electron correlation effect on the OEOP and TEOP transition energies and rates is shown in Fig.~\ref{fig1}. The correlation contributions to the transition energy from $\{n6l3\}$ are with respect to the MB. The correlation contribution to the transition energies from the $^{1}$P$_{1}$ upper level decrease smoothly, while increasing with Z for the others. The correlation effect contributes to the dipole allowed transition energy by 0.2 eV to 1.0 eV, while it is 0.2 eV to 1.5 eV for dipole forbidden transitions. However, for the TEOP transition, the contribution to the transition energy is 0.2 eV to 1.5 eV.
The percentage correlation contribution to the OEOP and TEOP transition energies from $\{n6l3\}$ with respect to the MB are given in Figs.~\ref{fig1}(a) and (b). The contribution from the electron correlation to the transition energy increases with increasing Z for TEOP transitions and the $^{3}$P$_{1}$-$^{1}$S$_{0}$ OEOP transition while the others decrease.

\begin{figure*}[!h]
\begin{minipage}[t]{1\linewidth}
\centering
\includegraphics[width=6in]{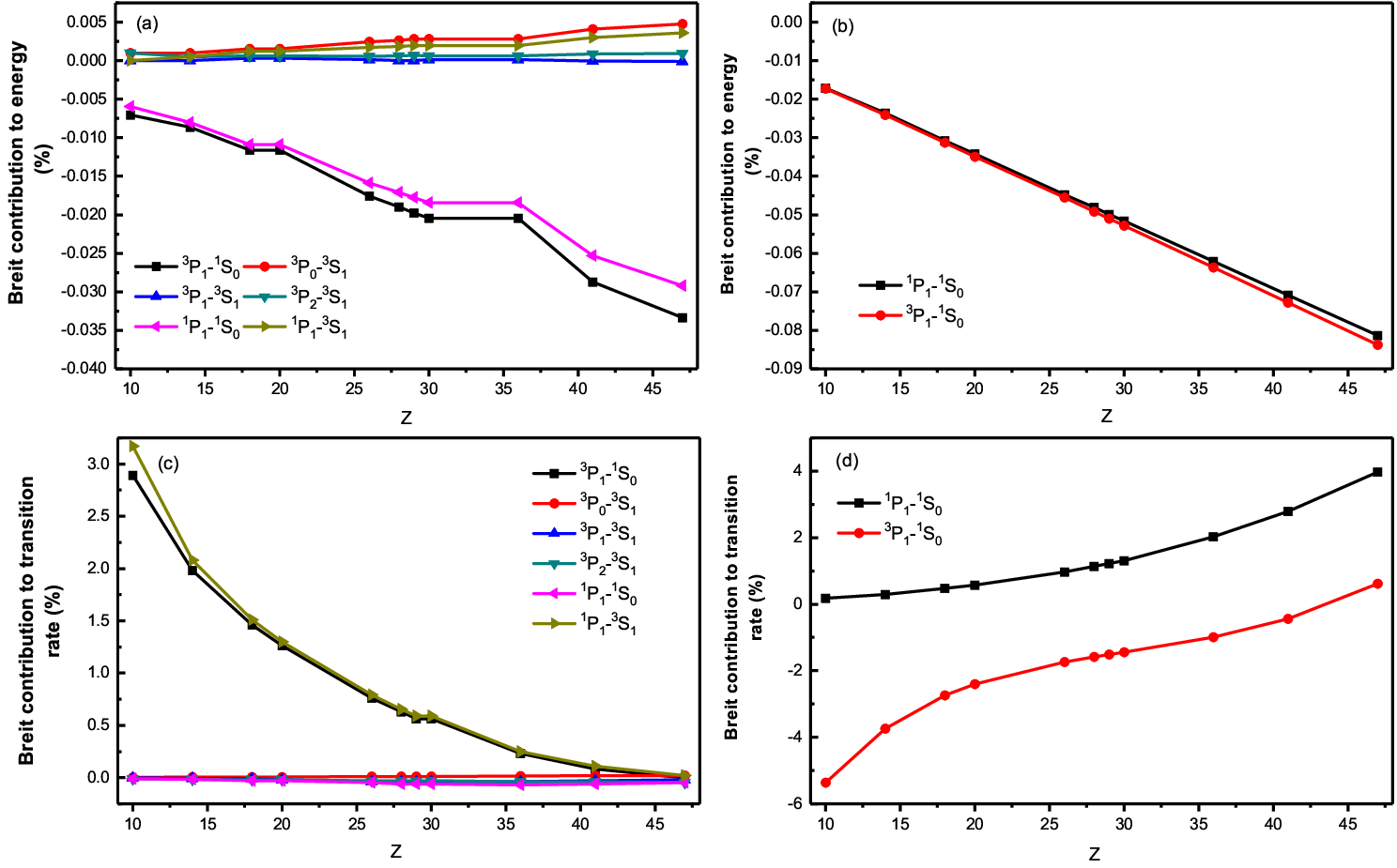}\label{Fig.2}
%\captionsetup{font={scriptsize}}   %
\centering
\caption{The Breit interaction effect on the transition energies and rates for OEOP and TEOP transitions in He-like ions. (a) The percentage Breit contribution to the OEOP transition energies of $2s2p-1s2s$. (b) The percentage Breit contribution to the TEOP transition energies of $2s2p-1s^{2}$. (c)The Breit contribution to the OEOP transition rates in the length gauge of $2s2p-1s2s$. (d)The Breit contribution to the TEOP transition rates in the length gauge of $2s2p-1s^{2}$. \label{fig2}}
\end{minipage}
\end{figure*}

Fig.~\ref{fig2} shows the contribution from the Breit interaction to the transition energies and rates of OEOP and TEOP transitions. It is seen in Fig.~\ref{fig2}(a) and (b) that the Breit interaction decreases the $^{1}$P$_{1}$-$^{1}$S$_{0}$ and $^{3}$P$_{1}$-$^{1}$S$_{0}$ transition energies for both OEOP and TEOP transitions, while it slightly increases the transition energy of the transition to the $^{3}$S$_{1}$ state in OEOP transitions. This is due to the Breit interaction reducing the binding energy of each state of the 2s2p configuration and also that of the $^{1}$S$_{0}$ state of the 1s2s configuration but slightly increasing the binding energy of the $^{3}$S$_{1}$ state of the 1s2s configuration, which makes the transition energy of $^{1}$P$_{1}$-$^{1}$S$_{0}$ and $^{3}$P$_{1}$-$^{1}$S$_{0}$ smaller and the energies of the other transitions to the $^{3}$S$_{1}$ state slightly increase. It is found that the contribution from the electron correlation is larger than that from the Breit interaction for low-Z elements, while the latter becomes significant for high-Z ions.

In Fig.~\ref{fig2}(c) and (d), the contributions of the Breit interaction to the transition rates in the length gauge of the OEOP and TEOP transitions is given. Unlike the correlation contribution, the Breit interaction reduces the rates of $^{1}$P$_{1}$-$^{1}$S$_{0}$ and $^{3}$P$_{1}$-$^{1}$S$_{0}$, and slightly increases the transition rates of other transitions to the $^{3}$S$_{1}$ state in OEOP processes. For TEOP transitions, the Breit interaction increases the transition rate of $^{1}$P$_{1}$-$^{1}$S$_{0}$ and $^{3}$P$_{1}$-$^{1}$S$_{0}$ with increasing Z. It can be seen from the figures that the Breit interaction contributions to the $^{1}$P$_{1}$-$^{3}$S$_{1}$ and $^{3}$P$_{1}$-$^{1}$S$_{0}$ OEOP transition rates are about $3.2\%$ and $2.8\%$ at Z = 10, respectively. These decrease with increasing Z, reaching approximately $0.1\%$ at Z = 47 for both transitions. However, for the TEOP transition, the Breit interaction contribution to the transition rate is about $0.5\%$-$5.5\%$ and $0.1\%$-$4\%$ for the $^{3}$P$_{1}$-$^{1}$S$_{0}$ and $^{1}$P$_{1}$-$^{1}$S$_{0}$ transitions, respectively. Since TEOP is a multi-electron process, the electron correlation effect plays an essential role in this transition, and the Breit interaction becomes more and more significant with increasing Z, as can be inferred from Fig.~\ref{fig1}(d) and \ref{fig2}(d).

The mixing of the CSFs leads to the feasibility of a TEOP transition that is strictly forbidden according to the selection rules. The main component of the $2s2p$ $^{1}$P$_{1}$ and $^{3}$P$_{1}$ states of the CSFs change from 67\% for Ne to 98\% for Ag, which indicates a change of the coupling scheme from $LSJ$ to $jj$ with a change in the nucleus and the interactions in these ions. The mixing from $1s2p$ $^{1}$P$_{1}$ and $^{3}$P$_{1}$ is tiny (less than $1\%$), even though it contributes to the main parts for the TEOP transitions. Because the $2p-1s$ resonance transition is strong, the TEOP transition matrix elements become non-zero because of this tiny mixing. Besides the mixing of the $1s2p$ with the excited states $2s2p$, there is also a  small mixing from $2s^2$, $2p^2$ contributing to the ground state $1s^2$ $^{1}$S$_{0}$.  Therefore, the $2p-2s$ and $2s-2p$ transition matrix elements could also contribute to the TEOP transition by mixing.

\begin{table}[h]
\footnotesize\rm
\centering
\caption{Transition energies (in eV) and rates (in sec$^{-1}$) in the length gauge of two-electron one-photon transitions from $2s2p$ to $1s^2$ in He-like ions. The numbers in the parentheses represent powers of ten.}
\label{Tab4}
\setlength{\tabcolsep}{8.0mm}{
%\scriptsize
\begin{tabular}{cccccccccc}
\hline
\hline
& &\multicolumn{2}{c}{$^{1}$P$_{1}$-$^{1}$S$_{0}$} &  \multicolumn{2}{c}{$^{3}$P$_{1}$-$^{1}$S$_{0}$ } \\
\multirow{1}{*}{Z}
          &                &  Energy   &  Rate      &  Energy   &  Rate                               \\
\midrule
   10	  &              & 1926.00	   & 6.030(9)	 & 1911.25   & 1.648(6)                           \\
    	  &Theory$^a$    & 1926.027   & 4.813(9)	 & 1911.507	 & 1.343(6)	                       \\
          &Theory$^b$    & 1928.844   & 1.269(10)   &           &                                  \\
   14     &              & 3844.42    & 1.232(10)   & 3822.79   & 1.955(7)\\
          &Theory$^a$    & 3843.901   & 9.391(9)    & 3822.746  & 1.715(7)\\
   18     &              & 6425.41   & 2.077(10)	 & 6396.07	 & 1.307(8)	                        \\
   	      &Theory$^a$    & 6424.385   & 1.568(10)   & 6396.059	 & 1.138(8)                       \\
          &Exp.$^c$      &            &             & 6390$^*$      &           \\
   20     &              & 7966.43    & 2.587(10)  & 7932.67     & 2.860(8)	\\
          &Theory$^a$    & 7967.522   & 1.945(10)	 & 7933.026	 & 2.491(8)	                   \\
          &Theory$^b$    & 7978.211   & 5.56(10)    &           &                       \\
   26     &              & 13604.75   & 4.346(10)   & 13553.46	 & 1.828(9)	\\
          &Theory$^a$    & 13604.082  & 3.342(10)	 & 13553.011 & 1.599(9)	   \\
   28     &               & 15827.11   & 5.001(10)	 & 15767.73	 & 2.955(9)      \\
   29     &              & 17003.44   & 5.338(10)	 & 16939.45	 & 3.676(9)                                   \\
   30     &              & 18223.52   & 5.682(10)	 & 18154.46	 & 4.512(9)                                    \\
   36     &             & 26475.88   & 7.897(10)	 & 26364.5	 & 1.220(10)                                    \\
   41     &              & 34603.61   & 1.002(11)	 & 34435.88	 & 2.232(10)                                        \\
   47     &             & 45918.8   & 1.313(11)	 & 45647.6	 & 3.917(10)                                         \\
\hline
\hline
\end{tabular}}
\begin{tablenotes}
\item[] $^{\rm a}$ Reference\cite{PhysRevA.84.062506}.
\item[] $^{\rm b}$ Reference\cite{Safronova1977}.
\item[] $^{\rm c}$ Reference\cite{Tawara2002b}.
\item[] $^*$ The observed experimental transition energy is about 6.39keV.\\
\end{tablenotes}
\end{table}


\section{Conclusion}

The energy levels, transition energies and transition rates for one- and two-electron radiative transitions from double K hole $2s2p$ to $1s2s$ and $1s^{2}$ configurations of He-like ions were calculated using MCDF method. A reasonable electron correlation model was constructed to study the electron correlation effects based on the active space. Breit interaction and QED effects were taken into account efficiently. The transition energies and rates were found to be in good agreement with those in previous work. It is emphasized in the present work that the TEOP transition is essentially caused by electron correlation effects. It is also found that the electron correlation effect and Breit interaction contribution to the transition energies of both OEOP and TEOP transitions decrease with increasing Z. Competition between the nucleus-electron Coulomb interaction and electron correlation was clearly found for lower Z ions. The former dominates in high Z ions. The calculated data will be helpful for future investigations on OEOP and TEOP transitions of He-like ions.

\section*{Acknowledgments}
This work was supported by National Nature Science Foundation of China, Grant No: U1832126, 11874051, National Key Research and Development Program of China, Grant No:2017YFA0402300.

%\begin{thebibliography}{10}
%\bibitem{book1} Goosens M, Rahtz S and Mittelbach F 1997 {\it The \LaTeX\ Graphics Companion\/}
%(Reading, MA: Addison-Wesley)
%\bibitem{eps} Reckdahl K 1997 {\it Using Imported Graphics in \LaTeX\ } (search CTAN for the file `epslatex.pdf')
%\end{thebibliography}
%%\bibliographystyle{myiopart-num}
\bibliographystyle{iopart-num}


%\bibliographystyle{myiopart-num}

\bibliography{me}
\end{CJK*}  %% end the Chinese environment
\end{document}